\documentclass [twocolumn,11pt]{article}

\usepackage{times}
\usepackage{graphicx} 

\begin{document}

\title{Secondary Muon Asymmetries at Sea Level with Low Systematics}
\author{J. Poirier \and C. D'Andrea \and M. Dunford}
\date{Center For Astrophysics at Notre Dame, Physics Dept., 
University of Notre Dame, Notre Dame, Indiana 46556 USA}



\maketitle

\begin{abstract}
Project GRAND has the capability of measuring the angle and identity
of single tracks of secondary muons at ground level.  
The array is comprised of 64 stations each
containing eight proportional wire planes with a 50 mm steel absorber
plate placed above the bottom two planes in each station.  The added
steel absorber plate allows muon tracks to be separated from the less
massive electrons.  Over 100 billion identified muon angles have
been measured.  With the high statistics available, it is possible to
obtain muon angular asymmetries with low systematics by subtracting 
west from east (and separately, south from north) angles; 
the subtraction eliminates most of the systematic errors  while still 
retaining adequately small statistical errors on the differences.  A
preliminary analysis is performed as a function of solar time to obtain the
effects on the muon rate due to effects of the sun.
\end{abstract}

\section{Introduction}

Cosmic rays are highly isotropic, yet small anisotropies do exist 
because of various effects such as the east-west asymmetry due to 
the magnetic field of the earth bending the predominately positively 
charged cosmic rays toward the east.  This effect has been observed 
some time ago; it is a constant effect depending upon the magnetic 
latitude of the detector and is independent 
of the time of day.  The value of this asymmetry 
is sensitive to the primary energy spectra and physical effects 
(such as, for example, the cosmic ray trajectory's deflection in 
the sun's or Earth's magnetic field) 
which depend on the low energy cut-off of the experiment.  

\section{The Compton-Getting Effect}

Another predicted asymmetry effect is that due to the 
Compton Getting Effect (CGE, see \cite{compton} and \cite{Cutler})
which describes a source of anisotropy due to the 
velocity of the detector relative to the frame in which the cosmic 
rays were created isotropically.  
The magnitude of the effect is given by:
\begin{equation}  
\frac{\Delta\alpha(\theta)}{\alpha}=[(2+\gamma)(v_0/c)]\cos \theta
\end {equation}
where $\Delta\alpha/\alpha$ is the fractional asymmetry, 
the quantity in square brackets is  $(F-A) / A$  with 
F the counting rate along the direction of the velocity and A the 
average rate, $v_0$ is the 
velocity of the detector relative to the rest frame of the cosmic rays, 
$\theta$ is the cosmic ray direction relative to $v_0$, cos $\theta$ 
is the projection of the cosmic ray along the forward direction 
of $v_0$, 
and $\gamma$ is the differential cosmic ray spectral index describing the 
energy spectrum of the primary cosmic ray.  

Using values of 30 km/sec for $v_0$ (the orbital speed of the Earth 
about the sun) and 2.7 for the spectral index,  
Equation 1 gives a CGE maximum amplitude of 0.47$\times10^{-3}$.  
The orbital speed of the Earth around the sun is used rather 
than the 220 km/sec velocity of 
the sun in the Galaxy since the data will be analyzed in a 
sun-centered frame and the larger effect of the Galactic speed will 
cancel out as the data are averaged over an integer number of solar years 
(assuming uniform data accumulation over this time).  

The asymmetry due to the Compton Getting effect using the orbital speed 
of the Earth around the sun will not be constant but depends upon the 
time of day relative to local noon (when the sun is due south of the 
experiment, which for GRAND is, on average, 12:45 EST).  
It is expected that the CGE asymmetry in E-W will be 
a positive maximum at midnight and negative maximum at noon (with a slight 
perturbation due to the Earth's rotational speed, which at GRAND is 
0.35~km/sec, about two 
orders smaller in magnitude than the Earth's speed of revolution about the sun).  
We study the east-west asymmetry (as well, the north-south) as a 
function of the solar time of day and subtract this CGE 
in order to measure any residual asymmetry due to other physical phenomena.   
  
\section{Experimental Array}

Project GRAND is an array of 64 proportional wire stations.  Each station 
is composed of four pairs of orthogonal planes.  Each plane contains 80 
detection wires; the planes are accurately aligned with the north/south or 
east/west directions.  
There is a steel plate above the bottom pair of planes; electrons scatter, 
stop, or shower because of the steel absorber plate but the higher mass 
muons are relatively unaffected.  
Because an electron 
will be misidentified as a muon approximately 4\% of the time and a muon 
will be misidentified as an electron also 4\% of the time, this arrangement 
allows Project GRAND to differentiate muon tracks with 96\% accuracy while 
retaining 96\% of them.  
Given the 80 channels of proportional wires in a plane and the vertical 
separation between the planes, Project GRAND is able to  
measure the direction of a muon track to 0.26$^\circ$, on average, in each 
of two orthogonal planes.  This geometrical arrangement of planes 
has a sensitivity cutoff-angle of $63^\circ$ from vertical.  
The muon threshold energy is 0.1~GeV for vertical 
tracks, increasing as 1/cos($\phi$) for $\phi$ inclined from vertical.  

Muons are identified and recorded at a rate of $\sim$2000 Hz.  Since January 
of 1998 through December of 2000, over 95  
billion muon tracks have been identified.   
  
Project GRAND measures the direction and identity of secondary muon tracks.   
In obtaining this east-west asymmetry by subtracting the west numbers 
from the east numbers (E-W), most systematic errors are eliminated. 
This E-W asymmetry is measured as well as 
a similar difference between muons originating from the south subtracted 
from the northern hemisphere (N-S).  
In taking these differences between the two hemispheres, angles 
of $0.0 \pm 0.7^\circ$ degrees from zenith have been deleted.   
These differences are then normalized to the sum.  
The high statistics available allows 
small differences to be seen with moderate error bars.

\section{Acceptance}

Since Project GRAND is not sensitive to the whole sky (its cutoff 
angle is 63$^\circ$ from zenith) and, like other ground-based detectors, 
it is more sensitive to cosmic rays coming from near its zenith, 
the appropriate data-average of cos $\theta$ must be obtained 
to use in Equation 1 in order to calculate the expected magnitude of 
the asymmetry in GRAND's data due to the CGE.  

The acceptance for Project GRAND, $Accept$, is given by: 
\begin{equation}
Accept=[1-0.537\tan\phi_x][1-0.537\tan\phi_y]\cos^3\phi
\end{equation}
The angle $\phi$ is the muon's angle from 
the vertical or zenith direction, $\phi_x$ is the projection of $\phi$ 
upon the xz-plane and $\phi_y$ is the yz-projection. 
It combines a geometrical 
factor, $[1-0.537\tan\phi_x][1-0.537\tan\phi_y]$, 
a cos $\phi$ factor describing the projection of the muons 
unto the zenith direction, and a factor, $cos^{2}\phi$, which describes 
the muon absorption 
in the Earth's atmosphere due to the increased path length in air for 
muons inclined from the 
vertical direction.   

The geometrical factor in Equation 1 arises from the 
arrangement of several horizontal proportional wire planes 
placed above each other in order to measure the angle of the muon from 
the vertical direction.  Since it is required for the muon to traverse 
both the top and bottom planes, as the muon's angle is inclined from 
zenith there is less area available in the planes 
for both planes to detect the muon.  These geometrical factors are the two 
[1 - 0.537tan] terms in Equation 2.  
The limiting projected angle is  
when a particle strikes the first wire of the top plane and the last wire 
of the bottom plane; this projected angle is $63^\circ$ from zenith.  
Any track of greater projected angle is not able to simultaneously intercept 
the top and bottom planes so will not be detected by this arrangement 
of proportional wire planes.  
The 0.537 in Equation 1 is the ratio between the horizontal  
width of the detector planes and the vertical spacing between the top 
and bottom planes of a detector station.  

The combined acceptance function is then 
folded with the $\cos \theta$ term in Equation 1 to find a 
data-averaged value for $\cos\theta$ to use in order to find 
the expected experimental asymmetry due to the CGE 
due to the orbital velocity of the Earth about the sun.  
Multiplying 
the CGE amplitude of 0.47$\times10^{-3}$ times the data-averaged 
cos $\theta$ (0.139) yields $0.065\times10^{-3}$ as the predicted Compton 
Getting effect 
which contributes to the 
east-west asymmetry as measured by Project GRAND; any residual 
asymmetry would then be due to other physical effects.  

\section{Data Analysis}

From January 1998 through December 2000,
673 data files have been created containing 
information on muon tracks.  These files have information on the local 
solar time, and direction of origin of the track.  From these data files 
only those files containing 
24 (or 48) continuous hours of data were selected.  
A smoothness test was applied 
to the data of each day to reduce the amount of error caused by 
possible temporary failures of a detector station or 
stoppages of the experiment.  These have the 
potential to produce 
structure in the subsequent analyses which would have nothing to do with the 
physics under study.

The data for each 24 hour period was 
divided into 48 half hour segments.  
Any 24 hour period where the standard deviation of the half hours was greater 
than 5\% of the average for that day was not used in the analysis.  
This left 304 days of data after this smoothness cut 
containing a total of 50 billion muon tracks for the subsequent analysis 
described below.  This cut is perhaps a little severe but helps insure that 
a possible residual 
asymmetry in a single day will not produce a noticeable effect in the 
sum over all the days; any remaining asymmetry should then be due to 
a physical effect other than the CGE.  

Since Project GRAND can determine the direction of incident muons, it was 
possible to separate muons from the eastern hemisphere from those of the 
western and an east-west (E-W) difference obtained.  
Also, muons from the southern hemisphere were subtracted from those 
from the northern hemisphere.
These E-W and N-S 
differences were obtained for each half-hour interval of data to 
obtain the corresponding asymmetry for that time period.  
These  
asymmetry numbers for each half hour of the solar day were summed; these sums 
were then 
normalized by dividing by the total number of muons accumulated during 
that half hour of the day.  The asymmetries were then 
plotted versus each half hour of the day in Eastern Standard Time (EST).  
The results are shown in Figures 1 and 2.  

\section{Fits}

Figure 1 displays the east-west asymmetry and Figure 2 the north-south.  
Equation 3 describes a curve with once- and twice-a-day  
variations which is fit to the data by finding the parameters which 
give a least-square error.  

\begin{center}
\begin {equation}
y=A+B\cos [15(x-C)] + D\cos [30(x-E)]
\end {equation}
\end{center}

The factor of 15 in the cosine function converts from hours 
to degrees in the once-a-day term.  
The factor of 30 in the second cosine function is 
twice as large; the additional 
factor of two gives the twice-per-day variation. 
$C$, $E$, and $x$ are hours in EST.  
The magnitude of $A$ is the average variation, 
$B$ and $D$ are parameters which describe the 
magnitude of the variations relative to the average, 
while $C$ and $E$ describe the locations of the peaks of these two 
variations in hour-of-day in EST.  
The values for these parameters are shown in Table 1.  For 
east-west, the twice-a-day variation is 82\% of the once-a-day variation.  
However, for north-south asymmetry, the twice-a-day variation 
is 61\% of the once-a-day value.  
The twice-a-day variation has the same strength for 
both the E-W and N-S asymmetries; 
the N-S once-a-day variation is 30\% larger 
than the E-W.

Fitting the E-W asymmetry data to 
Equation 3 yields the fitted parameters in Table 1.  
The errors shown assume no correlations and are, therefore, underestimates 
for those errors which are correlated.  
The largest asymmetry parameter is ``A" for the E-W subtracted data of 
$-3.35\times10^{-3}$.  This is due to the magnetic field of the 
Earth and the positive charge of the primary cosmic ray spectrum.  
The non-constant asymmetry is the  most negative at 18.1 hours and least 
negative at 11.2 hours with the value fluctuating 
between $-3.78\times10^{-3}$ and 
$-2.77\times10^{-3}$, respectively; the 
difference of these two extremes is $1.01\times10^{-3}$. 
The curve generated from fitting the 
N-S data to Equation 3 has an average 
value of $0.75\times10^{-3}$. 
The North-South asymmetry is the greatest at 8.9 hours and 
least at 16.7 hours yielding asymmetry values of $1.38\times10^{-3}$ 
and $0.17\times10^{-3}$, respectively; the difference of these two extremes 
is $1.21\times10^{-3}$, a larger difference than in the E-W case.  

The CGE effect predicts an east-west asymmetry value (the once-a-day B 
parameter) of $-0.065\times10^{-3}$ for the Project GRAND detectors.  
The remaining (now larger) asymmetry 
of $0.35\times10^{-3}$ is therefore be due to physics other than the CGE.  
One such possibility is the effect of the sun's interplanetary magnetic 
field deflecting the cosmic rays on their way toward the Earth.   
The CGE predicts no 
asymmetry for the north-south direction, so the $.43\times10^{-3}$ 
measured valule  
is due to other physical phenomena.   

\begin{table}[htbp]
\caption{Parameters of the fit coefficients in Equation 3}
\begin{center}
\begin{tabular}{ccc}\hline\hline
Parameter & E - W & N - S\\
\hline \hline
A & -0.00335 $\pm$ 0.00001 & 0.00075 $\pm$ 0.00001\\
\hline \hline
B & 0.00033 $\pm$ 0.00002 & 0.00044 $\pm$ 0.00002\\
\hline
D & 0.00028 $\pm$ 0.00002 & 0.00026 $\pm$ 0.00002\\
\hline \hline
C & 10.0 $\pm$ 0.2h & 7.0 $\pm$ 0.1h \\
\hline
E & 11.5 $\pm$ 0.1h & 9.7 $\pm$ 0.1 h \\
\hline \hline
\end{tabular}
\end{center}
\end{table}

\section{acknowledgemens}
Project GRAND is funded through the University of Notre Dame and private 
donations.

 \begin{figure*}[t]
\includegraphics*[width=11.0cm]{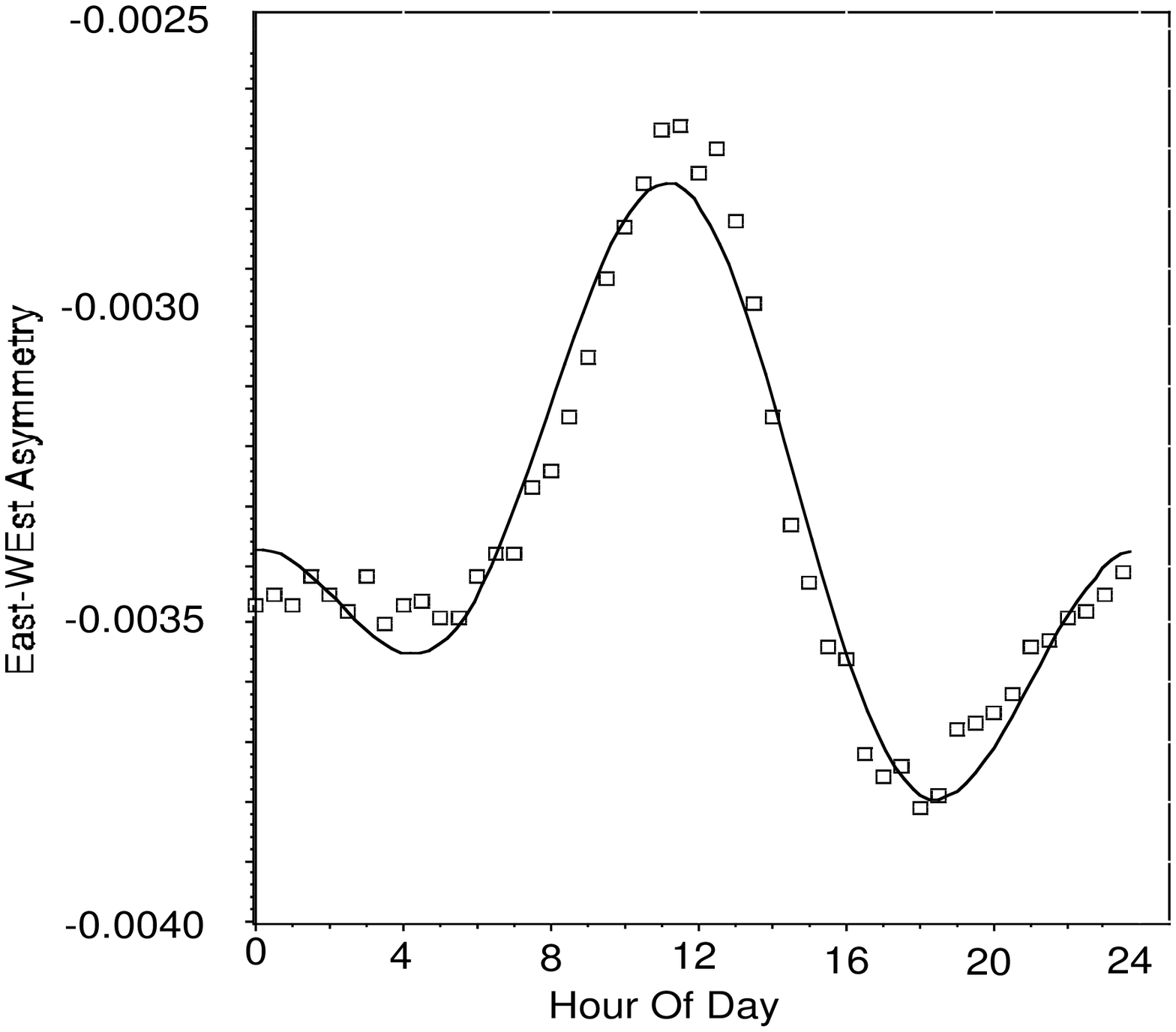} 

 \vspace{0mm}\caption{East - west asymmetry: Muons 
from the west subtracted from those from the east.  
The curve is the fit to Equation 1 
with the coefficients as listed in Table 1.  The abscissa is in Eastern 
Standard Time. Project Grand's local noon is, on average, 12:45 EST.}
 \end{figure*}


 \begin{figure*}[t]

\vspace{-40mm}
\includegraphics*[width=11.0cm]{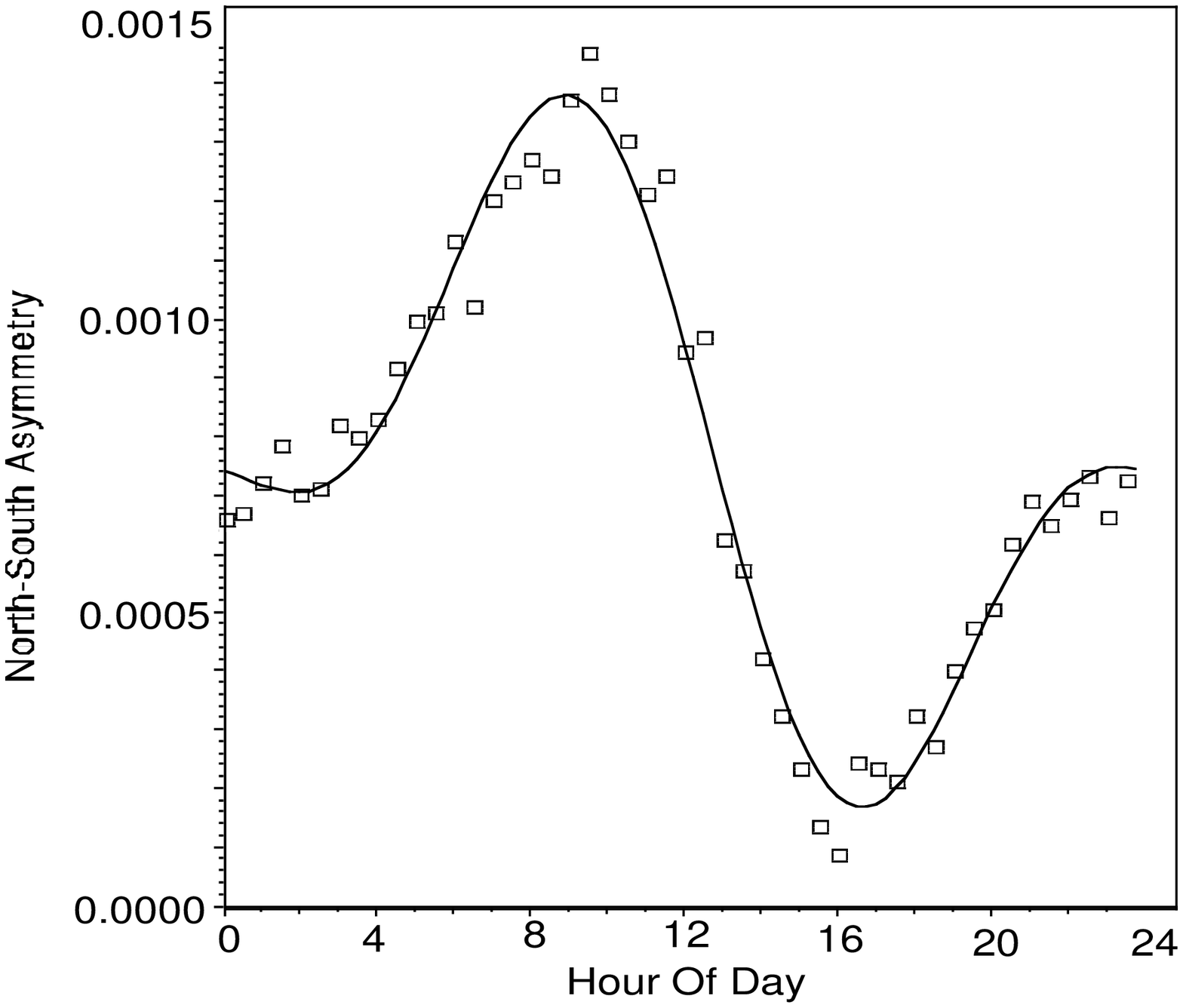}
\vspace{0mm}\caption{North - south asymmetry. Muons 
from the south subtracted from those from the north.  
The curve is the fit to Equation 1 
with the coefficients as listed in Table 1.  The abscissa is in Eastern 
Standard Time. Project Grand's local noon is, on average, 12:45 EST.}
 \end{figure*}

\end{document}